\documentclass[a4paper]{jpconf}
\usepackage{graphicx}

\newcommand{\ket}[1]{|#1\rangle}

\newcommand{\matrixel}[3]{\langle #1 | #2 | #3 \rangle}

\newcommand{\bfD}{{\mathbf D}}

\newcommand{\bfr}{{\mathbf r}}
\newcommand{\bfp}{{\mathbf p}}
\newcommand{\balpha}{{\mbox{\boldmath$\alpha$}}}
\newcommand{\bnabla}{{\mbox{\boldmath$\nabla$}}}
\newcommand{\veps}{\varepsilon}

\newcommand{\aZ}{\alpha Z}
\newcommand{\stgr}{{}^2P_{1/2}}
\newcommand{\stex}{{}^2P_{3/2}}

\newcommand{\lstgr}{[(1s)^2\,(2s)^2\,2p]\,{}^2P_{1/2}}
\newcommand{\lstex}{[(1s)^2\,(2s)^2\,2p]\,{}^2P_{3/2}}
\newcommand{\dgrec}{\Delta g_{\mathrm{rec}}}

\newcommand{\dgrecL}{\Delta g_{\mathrm{rec}}^{\mathrm{1-el,L}}}

\newcommand{\dgrecH}{\Delta g_{\mathrm{rec}}^{\mathrm{1-el,H}}}
\newcommand{\dgrecm}{\Delta g_{\mathrm{rec}}^{\mathrm{m-el}}}
\newcommand{\dgrecnr}{\Delta g_{\mathrm{rec}}^{\mathrm{non-rel}}}
\newcommand{\dgrecz}{\Delta g_{\mathrm{rec}}^{\mathrm{(1)}}}
\newcommand{\Tzero}{\left[ \bfr \times \balpha \right]_z}
\newcommand{\Wrec}{W_{\mathrm{rec}}}

\newcommand{\Hint}{H_{\mathrm{int}}}

\newcommand{\Vscr}{V_{\mathrm{scr}}}
\newcommand{\ArB}{{}^{40}\mathrm{Ar}^{13+}}
\begin{document}
\title{Nuclear recoil correction to the \emph{g} factor of boron-like argon}

\author{Arseniy A Shchepetnov$^{1,2}$, Dmitry~A~Glazov$^{1,2,3}$, Andrey~V~Volotka$^{1,3}$, Vladimir~M~Shabaev$^1$, Ilya~I~Tupitsyn$^1$ and G\"unter~Plunien$^3$}

\address{
$^1$ Department of Physics, St. Petersburg State University, Oulianovskaya 1, Petrodvorets, 198504 St. Petersburg, Russia
}
\address{
$^2$ State Scientific Centre ``Institute for Theoretical and Experimental Physics'' of National Research Centre ``Kurchatov Institute'', B. Cheremushkinskaya st. 25, 117218 Moscow, Russia
}
\address{
$^3$ Institut f\"ur Theoretische Physik, Technische Universit\"at Dresden, Mommsenstra{\ss}e~13, 01062 Dresden, Germany
}
\ead{d.glazov@spbu.ru}

\begin{abstract}
The nuclear recoil effect to the \emph{g} factor of boron-like ions is investigated. The one-photon-exchange correction to the nuclear recoil effect is calculated in the non-relativistic approximation for the nuclear recoil operator and in the Breit approximation for the interelectronic-interaction operator. The screening potential is employed to estimate the higher-order contributions. The updated \emph{g}-factor values are presented for the ground $\stgr$ and first excited $\stex$ states of B-like argon $\ArB$, which are presently being measured by the ARTEMIS group at GSI.

\end{abstract}

\section{Introduction}
During the last 15 years, the \emph{g}-factor measurements in low-$Z$ ions have reached an accuracy of $10^{-10}$ \cite{haefner:00:prl,verdu:04:prl,sturm:11:prl,sturm:13:pra,wagner:13:prl} and motivated corresponding theoretical investigations.
In particular, the most accurate value of the electron mass was obtained in these studies \cite{sturm:14:n}. The case of Li-like silicon manifests presently the most accurate verification of the many-electron QED effects in magnetic field \cite{wagner:13:prl,volotka:14:prl}. Experimental and theoretical investigations of the \emph{g} factor of heavy few-electron ions will provide stringent tests of bound-state QED in strong nuclear field. Moreover, they will serve for an independent determination of the fine structure constant, provided simultaneous investigations of H-like and B-like heavy ions of the same isotope will be performed \cite{shabaev:06:prl,volotka:14:prl-np}.

First measurement of the \emph{g} factor of a B-like highly charged ion sensitive to the QED effects was performed in Ref.~\cite{soriaorts:07:pra}. The ARTEMIS project presently implemented at GSI will use the laser-microwave double-resonance spectroscopy to measure with ppb accuracy the Zeeman splittings of both ground state and first excited state in B-like argon \cite{lindenfels:13:pra}. Corresponding theoretical predictions for the \emph{g} factor and the non-linear effects in magnetic field have been reported in Ref.~\cite{glazov:13:ps}. In this contribution we report on the recent progress for the nuclear recoil effect evaluated with more rigorous consideration of the screening correction. Namely, the contribution of the one-photon-exchange diagrams for the nuclear recoil effect has been calculated in the non-relativistic approximation. Total results for the \emph{g} factor of $\stgr$ and $\stex$ states of B-like argon presented here also include more accurate values of the interelectronic-interaction correction of the order $1/Z^2$ and higher.

The relativistic units ($\hbar=m=c=1$) and the Heaviside charge unit ($\alpha=e^2/(4\pi), e<0$) are used in the paper. Electron-to-nucleus mass ratio is written as $m/M$ for clarity.

\section{Nuclear recoil effect}

The theory of the nuclear recoil effect for the atomic \emph{g} factor to the leading orders in the parameter $\aZ$ was developed in a number of papers, see, e.g. Refs.~\cite{phillips:49:pr,faustov:70:plb,grotch:70:pra,grotch:71:pra,yelkhovsky:01,pachucki:08:pra} and references therein.
The rigorous QED theory valid to all orders in $\aZ$ and to first order in electron-to-nucleus mass ratio $m/M$ was developed in Ref.~\cite{shabaev:01:pra}. In Ref.~\cite{shabaev:02:prl} the corresponding numerical results were presented for $1s$ state.
Since the contributions of the second and higher orders in $m/M$ are negligible at the present level of accuracy, we do not consider them in the present paper.

First, we introduce the one-electron and the many-electron parts of the nuclear recoil correction, and present the former as the sum of the low-order and the higher-order terms:
\begin{eqnarray}
  \dgrec = \dgrecL + \dgrecH + \dgrecm
\,.
\end{eqnarray}
The lower-order term is given by the expression \cite{shabaev:01:pra}
\begin{eqnarray}
\label{eq:dgrecL}
  \dgrecL &=& \frac{1}{M_J} \frac{m}{M} \, \Bigg\{
    - \matrixel{a}{ \Big[ \bfr \times ( \bfp - \bfD(0) ) \Big]_z }{a} \phantom{\Bigg\}\,.}
\nonumber\\
& & + {\sum_n^{\veps_n\neq\veps_a}}
      \frac{\matrixel{a}{\Tzero}{n}}{\veps_a-\veps_n}
        \matrixel{n}{ \left[ \bfp^2 - \bfp \cdot \bfD(0) - \bfD(0) \cdot \bfp \right] }{a} \Bigg\}
\,.
\end{eqnarray}
Here $\ket{a}$ is the one-electron reference state, $M_J$ is the $z$-projection of the total angular momentum, while the $z$-axis is directed along the external magnetic field, $\balpha$ stands for the vector of Dirac matrices, $\bfp=-\imath\bnabla$ is the momentum operator, and
\begin{eqnarray}
  \bfD(\omega) = \aZ \left( \balpha \frac{\exp{(\imath\omega r)}}{r} + \bnabla (\balpha\cdot\bnabla) \frac{\exp{(\imath\omega r)}-1}{\omega^2 r} \right)
\,.
\end{eqnarray}

While $\dgrecL$ gives the one-electron nuclear recoil correction complete in orders $m/M (\aZ)^0$ and $m/M (\aZ)^2$, the higher-order term $\dgrecH$ contains contributions of the order $m/M (\aZ)^3$ and higher. Numerical evaluation of this term to all orders in $\alpha Z$ was done in \cite{shabaev:02:prl} for $1s$ state only. In this work, we estimate the uncertainty due to unknown value of $\dgrecH$ as $(\aZ)^3 \dgrecL$.

The many-electron part of the nuclear recoil correction for an atom with one electron over closed shells can be found from Eqs.~(73) and (92) of Ref.~\cite{shabaev:01:pra} employing the formalism of redefined vacuum \cite{shabaev:02:prep}. In this way, we derive the following expression:
\begin{eqnarray}
\label{eq:dgrecm}
  \dgrecm &=& \frac{2}{M_J} \frac{m}{M} \, \sum_c \Bigg\{
    \Big[ \matrixel{a}{\bfr}{c} \times \matrixel{c}{\big[ \bfp - \bfD(\Delta) \big]}{a} \Big]_z
\nonumber\\
& & - {\sum_n^{\veps_n\neq\veps_a}}
      \frac{\matrixel{a}{\Tzero}{n}}{\veps_a-\veps_n}
      \matrixel{n}{\Big[ \bfp - \bfD(\Delta) \Big]}{c} \cdot
      \matrixel{c}{\Big[ \bfp - \bfD(\Delta) \Big]}{a}
\nonumber\\
& & - {\sum_n^{\veps_n\neq\veps_c}}
      \frac{\matrixel{c}{\Tzero}{n}}{\veps_c-\veps_n}
      \matrixel{n}{\Big[ \bfp - \bfD(\Delta) \Big]}{a} \cdot
      \matrixel{a}{\Big[ \bfp - \bfD(\Delta) \Big]}{c}
\nonumber\\
& & + \matrixel{a}{ \bfD'(\Delta) }{c} \cdot
      \matrixel{c}{\Big[ \bfp - \bfD(\Delta) \Big]}{a}
    \Big( \matrixel{a}{\Tzero}{a} - \matrixel{c}{\Tzero}{c} \Big)
  \Bigg\}
\,.
\end{eqnarray}
Here the summation over $\ket{c}$ runs over all closed-shell electrons, $\bfD'(\omega)=\partial \bfD(\omega) / \partial \omega$, and $\Delta=\veps_a-\veps_c$.

We calculate the contributions $\dgrecL$ and $\dgrecm$ according to Eqs.~(\ref{eq:dgrecL}) and (\ref{eq:dgrecm}) for $\stgr$ and $\stex$ states. The numerical computation is performed employing the standard algebra for angular coefficients and the dual kinetic balance approach \cite{shabaev:04:prl} to construct the finite basis set of the radial functions.
Apart from the Coulomb potential (with account for the finite nuclear size), we use the core-Hartree and Kohn-Sham screening potentials. The explicit expressions for this potentials can be found e.g. in~\cite{cowan}, while the examples of their numerical implementations and applications in various atomic structure calculations can be found, e.g. in Refs.~\cite{volotka:14:prl,sapirstein:01:pra,glazov:06:pla,yerokhin:07:pra,artemyev:07:prl}. The results for B-like argon are presented in the first and second lines of Table~\ref{tab:dgrec}.

\begin{figure}
\includegraphics{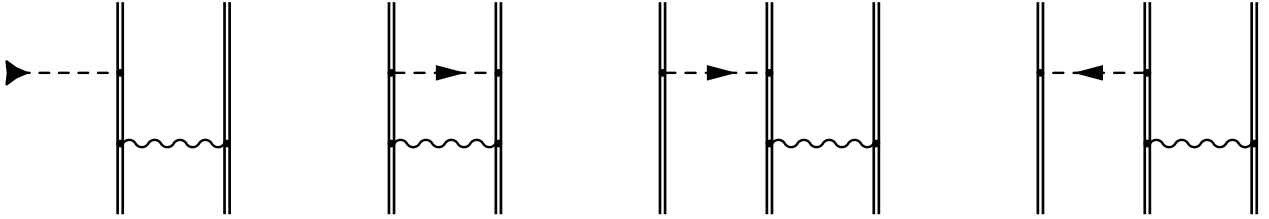}
\caption[]{Diagrams corresponding to $\dgrecz$ contribution to the \emph{g} factor, given by Eq.~(\ref{eq:dgrecz}): the first-order interelectronic-interaction correction to the nuclear recoil effect. The dashed lines with the triangle and with the arrow correspond to the one-electron and many-electron parts of the non-relativistic recoil operator $\Wrec$ (\ref{eq:Wrec}).
\label{fig:dgrecz}}
\end{figure}

\begin{figure}
\begin{center}
\includegraphics{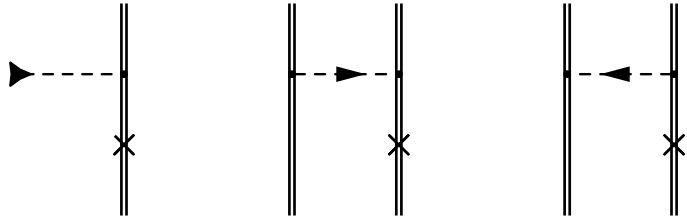}
\end{center}
\caption{Additional counter-term diagrams for the screening potential.
\label{fig:dgrecV}}
\end{figure}

The next step in our consideration is to take into account the interelectronic-interaction beyond the screening-potential approximation, namely, to calculate the first-order correction to the nuclear recoil effect within the perturbation theory. In the non-relativistic limit Eqs.~(\ref{eq:dgrecL}) and (\ref{eq:dgrecm}) yield together the well-known expression \cite{phillips:49:pr}
\begin{eqnarray}
\label{eq:dgrecnr}
  \dgrecnr = \frac{1}{M_J} \, \matrixel{A}{\Wrec}{A}
\,,\\
\label{eq:Wrec}
  \Wrec = \frac{m}{M} \sum_{j,k}[\bfr_j \times \bfp_k]_z
\,.
\end{eqnarray}
Here $\ket{A}$ is the reference-state many-electron wave function in the non-interacting-electrons approximation, i.e. the Slater determinant for $(1s)^2\,(2s)^2\,2p_j$ configuration. We employ the approximation given by Eqs.~(\ref{eq:dgrecnr}),~(\ref{eq:Wrec}) to evaluate the first-order interelectronic-interaction correction to the nuclear recoil effect. The interaction operator is taken in the Breit approximation,
\begin{eqnarray}
\label{eq:Hint}
  \Hint = \alpha \sum_{j<k}
    \left( \frac{1-\balpha_j\cdot\balpha_k}{r_{jk}}
         - \frac{1}{2} \left[ (\balpha_j\cdot\bnabla_j),
           \left[ (\balpha_k\cdot\bnabla_k), r_{jk} \right]
           \right] \right)
\,.
\end{eqnarray}
The general expression for this contribution is
\begin{eqnarray}
\label{eq:dgrecz}
  \dgrecz = \frac{2}{M_J} \, \sum_N^{E_N \neq E_A}
    \frac{\matrixel{A}{\Wrec}{N}\matrixel{N}{\Hint}{A}}{E_A-E_N}
\,,
\end{eqnarray}
where the summation runs over the complete spectrum of the many-electron states $\ket{N}$, constructed as the Slater determinants from the one-electron solutions of the Dirac equation. Substitution of the two-electron operators $\Wrec$ and $\Hint$ leads to the rather lengthy formulae, which are not presented here, therefore. The structure of these formulae can be easily understood from the corresponding diagrams depicted in Fig.~\ref{fig:dgrecz}. The first diagram corresponds to the ``one-electron'' part of $\Wrec$, i.e. to the case of $j=k$ in Eq.~(\ref{eq:Wrec}). Other diagrams correspond to the ``two-electron'' part of $\Wrec$, i.e. to the case of $j\neq k$.
We calculate $\dgrecz$ numerically for the pure Coulomb potential, as well as for the core-Hartree and Kohn-Sham screening potentials. When the screening potential is included in zeroth-order in the Dirac equation, the corresponding counter-term shall be taken into account for the first-order correction $\dgrecz$. It can be written as the following replacement in Eq.~(\ref{eq:dgrecz}),
\begin{equation}
  \Hint \to \Hint -
  \sum_j \Vscr(r_j)
\,.
\end{equation}
The diagrams for the counter-term are shown in Fig.~\ref{fig:dgrecV}.

In Table~\ref{tab:dgrec} the terms $\dgrecL$, $\dgrecm$, and $\dgrecz$ are presented for the case of B-like argon $\ArB$ for Coulomb, core-Hartree and Kohn-Sham potentials. The Kohn-Sham value is taken for the final result, while the uncertainty is estimated as the difference between the values for Coulomb and Kohn-Sham potentials. This rather conservative estimation, i.e. 100\% of the contribution of higher orders, is based on the observation, that the effect of the screening potential for $\dgrecL$ and $\dgrecm$ constitutes only 50\% of the total interelectronic-interaction correction obtained. Another minor uncertainty is related to $\dgrecH$, its estimation ($(\aZ)^3 \dgrecL$) amounts to $0.041$ and $0.021$ for $\stgr$ and $\stex$ states, respectively. In comparison to the previously published values \cite{glazov:13:ps}, the obtained accuracy of the nuclear recoil effect is 2 times better for $\stgr$ state and 4 times better for $\stex$ state.

\begin{table}
\caption{Individual contributions to the nuclear recoil correction to the \emph{g} factor of B-like argon $\ArB$ ($m/M=13.7308\cdot10^{-6}$) for $\stgr$ and $\stex$ states. The units are $10^{-6}$.
\label{tab:dgrec}}
\vspace{0.5cm}
\begin{tabular}{|c||r|r|r||r|r|r|}
\hline
& \multicolumn{3}{c||}{$^{2}P_{1/2}$} & \multicolumn{3}{|c|}{$^{2}P_{3/2}$}\\
\hline
&   Coulomb & core-Hartree & Kohn-Sham &   Coulomb & core-Hartree & Kohn-Sham \\
\hline
$\dgrecL$ &$  -18.208$&$     -18.223$&$  -18.221$&$   -9.099$&$ -9.109$&$ -9.108$\\
%
%
$\dgrecm$ &$    7.548$&$       8.218$&$    8.332$&$    3.698$&$   4.064$&$  4.119$\\
%
%
$\dgrecz$ &$    1.434$&$       0.929$&$    0.790$&$    0.704$&$ 0.459$&$  0.387$\\
\hline
sum  &$   -9.226$&$      -9.076$&$   -9.099$&$   -4.697$&$   -4.586$&$  -4.602$\\
\hline
$\dgrec$  &\multicolumn{3}{r||}{$-9.10(13)$}&\multicolumn{3}{r|}{$  -4.60(9)$}\\
\hline
\end{tabular}
\end{table}

\section{\emph{g} factor of boron-like argon} 

Table \ref{tab:g-th} represents the individual contributions to the \emph{g} factors of B-like argon for the ground $\lstgr$ and first excited $\lstex$ states. As compared to the previous compilation \cite{glazov:13:ps}, two terms are improved: the interelectronic interaction of the second and higher orders ($1/Z^{2+}$) and the nuclear recoil effect. Evaluation of the latter has been presented in the previous section. The higher-order interelectronic-interaction correction was calculated in \cite{glazov:13:ps} for the ground state within the Breit approximation employing the large-scale configuration interaction method with the Dirac-Fock and Dirac-Fock-Sturm basis functions (CI-DFS). The 10\% uncertainty was ascribed to it due to the moderate basis employed in the calculations and relatively poor convergence of the result with respect to the basis size. Recently we have performed calculations with larger basis and found justification for 2 times smaller uncertainty of the result. It is supported by the independent calculation of the $1/Z^2$-term performed within the perturbation theory. The corresponding calculation for the $\stex$ state yields the new value for this contribution, while the all-order CI-DFS result is still in demand. The details on this calculation will be published elsewhere. In total, we have an improvement by a factor of $1.5$ for the \emph{g} factor of the $\stgr$ state.

\begin{table}
\caption{Individual contributions to the \emph{g} factor of boron-like argon for $\stgr$ and $\stex$ states.
\label{tab:g-th}}
\vspace{0.5cm}
\begin{tabular}{|ll|r@{}l|r@{}l|}
\hline
&& \multicolumn{2}{c|}{$\stgr$}
& \multicolumn{2}{c|}{$\stex$}\\
\hline
Dirac value                           &&    0.&663 775 447  &    1.&331 030 389 \\
Finite nuclear size                   &&    0.&000 000 000  &    0.&000 000 000 \\
One-photon exchange  & $\sim 1/Z$      &    0.&000 657 525  &    0.&000 481 188 \\
Many-photon exchange & $\sim 1/Z^{2+}$ & $-$0.&000 007 5 (4)&    $-$0.&000 003 (3) \\
One-loop QED     & $\sim \alpha$       & $-$0.&000 769 9 (5)&    0.&000 779 6 (8) \\
Higher-order QED & $\sim \alpha^{2+}$  &    0.&000 001 2 (1)& $-$0.&000 001 2 (1) \\
Nuclear recoil                        && $-$0.&000 009 1 (2)& $-$0.&000 004 6 (1) \\
\hline
Total                                 &&    0.&663 647 7 (7)&    1.&332 282 (3) \\
\hline
\end{tabular}
\end{table}

\ack

The work was supported in part by~DFG (Grant No. VO~1707/1-2), by~GSI, by~RFBR (Grants No. 14-02-31316 and 13-02-00630), by SPbSU (Grants No. 11.38.269.2014, and No. 11.38.261.2014), and by~the~Helmholtz-Rosatom grant provided~via~FAIR--Russia Research Center.

\end{document}